\documentclass[a4paper,fleqn,usenatbib]{mnras}

\usepackage{newtxtext,newtxmath}

\usepackage[T1]{fontenc}
\usepackage{ae,aecompl}

\usepackage{graphicx}
\usepackage{epsfig}
\usepackage{amsfonts}
\usepackage{subfigure}

\usepackage{graphicx}	
\usepackage{amsmath}	
\usepackage{amssymb}	

\usepackage{epsfig}
\usepackage{subfigure}

\title[Structure formation in clustering DBI ...]{Structure formation in clustering DBI dark energy model with constant sound speed}

\author[K. Fahimi et al.]{
K. Fahimi,$^{1}$\thanks{fahimi0678@gmail.com}
K. Karami,$^{1}$\thanks{kkarami@uok.ac.ir}
S. Asadzadeh,$^{2}$\thanks{ss.asadzadeh@gmail.com}
and K. Rezazadeh$^{1}$\thanks{rezazadeh86@gmail.com}
\\
$^{1}$Department of Physics, University of Kurdistan, Pasdaran Street, P.O. Box 66177-15175, Sanandaj, Iran\\
$^{2}$Research Institute for Astronomy and Astrophysics of Maragha (RIAAM), P.O. Box 55134-441, Maragha, Iran
}

\date{Accepted XXX. Received YYY; in original form ZZZ}

\pubyear{2018}

\begin{document}
\label{firstpage}
\pagerange{\pageref{firstpage}--\pageref{lastpage}}
\maketitle

\begin{abstract}

Within the framework of DBI non-canonical scalar field model of dark energy, we study the growth of dark matter perturbations in both the linear and non-linear regimes. In our DBI model, we consider the anti-de Sitter warp factor $f(\phi)=f_0\, \phi^{-4}$ with constant $f_0>0$ and assume the DBI dark energy to be clustered and its sound speed $c_s$ to be constant. In the linear regime, we use the Pseudo-Newtonian formalism to obtain the growth factor of dark matter perturbations and conclude that for smaller $c_s$ (or $\tilde{f_0} \equiv f_0 H_0^2/M_P^2$), the growth factor of dark matter is smaller for clustering DBI model compared to the homogeneous one. In the non-linear regime based on the spherical collapse model, we obtain the linear overdensity $\delta_c(z_c)$, the virial overdensity $\Delta_{\rm vir}(z_c)$, overdensity at the turn around $\zeta(z_c)$ and the rate of expansion of collapsed region $h_{\rm ta}(z)$. We point out that for the smaller $c_s$ (or $\tilde{f_0}$), the values of $\delta_c(z_c)$, $\Delta_{\rm vir}(z_c)$, $\zeta(z_c)$ and $h_{\rm ta}(z)$ in non-clustering DBI models deviate more than the $\Lambda$CDM compared to the clustering DBI models. Finally, with the help of spherical collapse parameters we calculate the relative number density of halo objects above a given mass and conclude that the differences between clustering and homogeneous DBI models are more pronounced for the higher-mass halos at high redshift.

\end{abstract}




\section{Introduction}
\label{section:introduction}

Observational data from various sources including the type Ia supernovae (SNeIa) \citep{Riess1998, Perlmutter1999, Kowalski2008}, cosmic microwave background (CMB) \citep{Komatsu2009, Jarosik2011, Komatsu2011, Planck2016}, large-scale structure (LSS), baryonic acoustic oscillations (BAO) \citep{Tegmark2004, Cole2005, Eisenstein2005, Percival2010, Blake2011, Reid2012}, high redshift galaxies \citep{Alcaniz2004}, high redshift galaxy clusters \citep{Wang1998, Allen2004}, and weak gravitational lensing \citep{Benjamin2007, Amendola2008, Fu2008} indicate that our present Universe is in an accelerating phase of expansion. In order to explain this cosmic acceleration in the framework of Einstein's general relativity (GR), we need to invoke an exotic form of matter-energy with negative pressure, the so-called dark energy (DE). One candidate for the DE is the cosmological constant $\Lambda$ \citep{Peebles2003}. However, the cosmological constant suffers from fundamental problems such as the fine-tuning and cosmic coincidence problems \citep{Weinberg1989, Sahni2000, Carroll2001, Padmanabhan2003, Copeland2006}. Because of these problems, the scientific community suggest dynamical candidates for the DE. The most important class of these models, for example, are the quintessence \citep{Caldwell1998, Erickson2002}, phantom \citep{Caldwell2002, Caldwell2003, Elizalde2004}, $k$-essence \citep{Chiba2000, Armendariz-Picon2000, Armendariz-Picon2001}, holographic \citep{Hovrava2000, Thomas2002}, chaplygin gas \citep{Kamenshchik2001}, generalized chaplygin gas \citep{Bento2002}, dilaton \citep{Gasperini2001, Arkani-Hamed2004, Piazza2004}, agegraphic \citep{Cai2007}, new agegraphic \citep{Wei2008}, and quintom \citep{Wei2005}.

Alternatively, the current accelerated expansion of the Universe can be described in the framework of extended theories of gravity, which is so-called modified gravity (MG) \citep{Ida2000, Brax2004, Atazadeh2006, Myrzakulov2011, Clifton2012, Guo2013, Mukherjee2014, Cai2016, Novikov2016, Novikov2016-2}. In the MG models, the gravitational part of the effective action is modified and it is no longer the one related to the Einstein term.

Indeed, the DE (or MG) not only affects the dynamics of the background cosmology through the modification of the Hubble parameter, but also changes the rates of formation and growth of collapsed structures (halos) via its fluctuations. Therefore, in addition to the expansion history, measuring the growth of structure can be used as a useful tool to distinguish between DE (or MG) models. Although different models of DE (or MG) may predict the same background very close to one in the $\Lambda$CDM model, they rarely ever create the same perturbations. Thus studying the large-scale structure provides valuable information about the nature of DE (or MG) \citep{Tegmark2004, Tegmark2006}. So far, the issue of structure formation in the context of DE models has been extensively considered in the literature \citep{Linder2003, Abramo2007, Sefusatti2011, Batista2013, Roupas2014,  Malekjani2015, Naderi2015, Nazari-Pooya2016, Batista2017, Rezaei2017, Rezaei2017-2}. Also, this subject has been widely studied in the framework of MG theories \citep{Koyama2006, Brax2012, Nesseris2013, Asadzadeh2016, Nunes2018}.

It is believed that the large-scale structures in the Universe result from the gravitational collapse of primordial small density perturbations \citep{Gunn1972, Press1974, White1978, Peebles1993, Peacock1999, Peebles2003, Ciardi2005, Bromm2011}. The initial seeds of the primordial perturbations are generated during the inflationary phase of our Universe \citep{Guth1981, Linde1990}. At early times of expansion, linear theory of perturbations is valid because the overdensities are small, and also the scales of interest in cosmology are much smaller than the Hubble horizon and the velocities are non-relativistic. Therefore, we can implement the Pseudo-Newtonian gravity that can be used to investigate the evolution of the density fluctuations in the linear regime. In the Pseudo-Newtonian formalism, the relativistic contributions appear as pressure terms in the Poisson equation, and in this way we can use the Newtonian hydrodynamical equations in an expanding Universe \citep{Abramo2008, Abramo2009JCAP}. It should be noted that at late times the perturbations grow so that the linear regime of perturbations is no longer valid, and we have to examine the evolution of overdensities in the non-linear regime. To this aim, it is convenient to follow the spherical collapse model (SCM) \citep{Gunn1972}, which is a simple analytical approach to study the non-linear perturbations. Investigation of the evolution of the fluctuations at this level provides some cosmological results which are observationally valuable. An interesting possibility in the context of SCM is that DE may mutate into a fluid with clustering properties similar to those of dark matter (DM) \citep{Abramo2008}. This effect is a generic feature of DE and originates from a simple idea: when pressure perturbations are large, the effective equation of state inside a collapsed region can be different from the one of its homogeneous component \citep{Abramo2008}. In \citet{Abramo2009JCAP}, it has been discussed that the clustering DE can show a detectable impact on the cosmological observables.

Notice that within the framework of canonical scalar fields like the quintessence model \citep{Caldwell1998, Erickson2002}, the perturbations of DE on small scales of structures are several orders of magnitude smaller than those in the DM and consequently the DE perturbations are usually neglected on these scales. This is because of the fact that the effective sound speed of quintessence perturbations is very close to the light speed ($c_{\rm eff}\sim 1$) which yields the growth of scalar field perturbations to be suppressed inside the sound horizon. However, in the non-canonical scalar fields like the $k$-essence models \citep{Armendariz-Picon2000, Chiba2000, Armendariz-Picon2001}, the DE perturbations can propagate with an effective sound speed much smaller than one and consequently the DE can be clustered in the $k$-essence models on sub-horizon scales.

In this paper, we focus on the study of structure formation in the context of Dirac-Born-Infeld (DBI) scalar field. The DBI scalar field is included in the class of non-canonical scalar fields whose kinetic energy in the action is different from the canonical one. The DBI scalar field has well-based motivations from the string theory. In fact the DBI scalar field is used to describe the dynamics of the radial coordinate of a D3-brane moving in a warped region (throat) of a compactification space \citep{Alishahiha2004, Silverstein2004}. In this description, the brane can be imagined as a point-like object whose speed is limited by the warp factor of the throat. Due to this speed limit, we usually introduce a parameter $\gamma$ which is analogous to the Lorentz factor in the special relativity. Note that in the case $\gamma>1$, the sound speed $c_s=1/\gamma$ of the DBI dark energy model is less than the light speed. The sound speed determines the velocity of perturbations propagation among the space, and therefore it is expected this parameter possesses a crucial role in the study of cosmological structure formation. In this paper, our main aim is to investigate the cosmological implications of the clustering ($c_{\rm eff}=0$) and non-clustering ($c_{\rm eff}=1$) DBI dark energies with the constant sound speed $c_s$ in both the linear and non-linear regimes of perturbations.

The structure of this paper is as follows. In Sec. \ref{section:background}, we study the background cosmology in the presence of DBI dark energy model. In Sec. \ref{section:linear}, we investigate the growth of linear perturbations in this model. In Sec. \ref{section:spherical_collapse}, we study the non-linear evolution of DM and DE fluctuations in the SCM. Finally, Sec. \ref{section:conclusions} is devoted to our conclusions.

\section{Background cosmology in DBI model}
\label{section:background}

In the framework of Einstein gravity, the action of DBI non-canonical scalar field is given by \citep{Silverstein2004}
\begin{equation}
\label{S}
S=\int d^{4}x\sqrt{-g}\left[\frac{M_P^2}{2}R+\mathcal{L}(X,\phi)\right]+S_m,
\end{equation}
where $M_{P}=(8\pi G)^{-1/2}$, $g$ and $R$ are the Planck reduced mass, the determinant of the background metric $g_{\mu\nu}$ and the Ricci scalar, respectively. Also $S_m$ is the action of matter field and $\mathcal{L}(X,\phi)$ is the Lagrangian density of the DBI scalar field defined as follows
\begin{equation}
\label{{mathcal}{L}}
\mathcal{L}(X,\phi)\equiv f^{-1}(\phi)\left[1-\sqrt{1-2f(\phi)X}\right]-V(\phi),
\end{equation}
where $f(\phi)$ is the warp factor. Besides, $X\equiv - g^{\mu \nu} \partial_\mu \phi~ \partial _\nu \phi/2$ and $V(\phi)$ are the canonical kinetic term and potential of the DBI scalar field $\phi$, respectively.

The energy density and pressure of the non-canonical scalar field DE $\phi$ corresponding to the DBI Lagrangian (\ref{{mathcal}{L}}) take the forms
\begin{align}
\label{rho_DBI}
\rho_d &\equiv 2X\mathcal{L}_{,\rm X} -\mathcal{L}=\frac{\gamma-1}{f(\phi)}+V(\phi),
\\
\label{p_DBI}
p_d &\equiv \mathcal{L}=\frac{\gamma-1}{\gamma f(\phi)}-V(\phi),
\end{align}
where the subscript ``$,X$'' denotes the partial derivative with respect to $X$. In addition, the parameter $\gamma$ is defined in analogy of the Lorentz boost factor as
\begin{equation}
\label{{gamma}}
\gamma\equiv\frac{1}{\sqrt{1-2f(\phi)X}},
\end{equation}
which determines the relativistic limit of brane motion in a warped background.

For a spatially flat Friedmann-Robertson-Walker (FRW) Universe containing the DBI dark energy and pressureless DM, variation of the action (\ref{S}) with respect to the metric reduces to the first and second Friedmann equations as follows
\begin{align}
 & H^{2}=\frac{1}{3M_{P}^{2}}(\rho_d+\rho_{m})=\frac{1}{3M_{P}^{2}}\left(\frac{\gamma-1}{f(\phi)}+V(\phi)+\rho_{m}\right),
 \label{H2}
 \\
 & \dot{H}=\frac{-1}{2M_{P}^{2}}\big(\gamma\dot{\phi}^{2}+\rho_{m}\big),
 \label{Hdot}
\end{align}
where $H=\dot{a}/a$ is the Hubble parameter with the scale factor $a(t)$. Note that for the flat FRW matric, the canonical kinetic term reads $X=\dot{\phi}^2/2$.

Taking the variation of action (\ref{S}) with respect to $\phi$ yields the equation of motion of the scalar field as
\begin{equation}\label{phidot}
  \ddot{\phi}+\frac{3 f_{,\phi}}{2 f}\dot{\phi}^2-\frac{f_{,\phi}}{f^2}+\frac{3H}{\gamma^2}\dot{\phi}+\left(V_{,\phi}+\frac{f_{,\phi}}{f^2}\right)\frac{1}{\gamma^3}=0,
\end{equation}
where the subscript ``$,\phi$'' indicates the partial differentiation with respect to $\phi$. This equation of motion can also be obtained from the continuity equation governing the scalar field
\begin{equation}
\label{{dot}{{rho}}_{phi}}
\dot{\rho}_d+3H(\rho_d+p_d)=0.
\end{equation}
Also the pressureless DM satisfies the following continuity equation
\begin{equation}\label{rhom}
 \dot{\rho}_m+3H\rho_m=0.
\end{equation}
This gives the solution
\begin{equation}
\rho_m=\rho_{m_0}a^{-3},
\end{equation}
where $\rho_{m_0}$ is the matter energy density at the present scale factor $a_0=1$.

Within the framework of non-canonical scalar field, the sound speed characterizing the propagation speed of the DE fluctuations $\delta\phi$ relative to the homogeneous background is defined as
\begin{equation}
\label{c_s,definition}
c^2_s \equiv \frac{p_{d,{\rm X}}}{\rho_{d,{\rm X}}}.
\end{equation}
Note that due to having a real and subluminal sound speed, it should satisfy the condition $0<c_{s}^{2}\leq1$. Using Eqs. (\ref{rho_DBI}) and (\ref{p_DBI}) in (\ref{{gamma}}), the sound speed in DBI dark energy reads
\begin{equation}
\label{c_s}
c_{s}=\sqrt{1-2f(\phi)X}=\frac{1}{\gamma}.
\end{equation}
For the case $c_s=1$ (or $\gamma=1$), from Eqs. (\ref{rho_DBI}) and (\ref{p_DBI}) we have $p_d=-\rho_d$ and consequently our DBI model is transformed to the $\Lambda$CDM one.

Note that the set of equations (\ref{H2}), (\ref{Hdot}), (\ref{phidot}) and (\ref{rhom}) are not independent of each other. Taking the time
derivative of Eq. (\ref{H2}) and using Eqs. (\ref{phidot}) and (\ref{rhom}), one can
obtain the second Friedmann equation (\ref{Hdot}). In what follows,
we take the set of Eqs. (\ref{H2}), (\ref{Hdot}), and (\ref{rhom}), which can
uniquely determine the dynamics of the Universe. Now to obtain the evolutionary behaviors of $\phi(a)$ and $H(a)$, we consider the anti-de Sitter (AdS) warp factor $f(\phi)=f_0\, \phi^{-4}$ with constant $f_0>0$ and following \citet{Spalinski2008, Tsujikawa2013, Amani2018}, we assume the sound speed $c_s$ to be constant. Although we assume the sound speed of the DBI dark energy to be constant, this does not confine considerably the generality of our discussion. Since  the present Universe experiences an intensive accelerating expansion, following \citet{Garriga1999}, we can regard the variation of the sound speed $c_s$ to be much slower than the increase rate of the cosmic scale factor $a$. In spite of the fact that an exact expression has been provided for the sound speed in Eq. (\ref{c_s}), but this quantity is slow-varying versus time, and it can be taken as a constant parameter to a good approximation. As we will see, one important consequence of this approximation is that, having the function of the warp factor $f(\phi)$, we can determine the function of the scalar potential $V(\phi)$, that we have no robust theoretical or experimental idea for it yet.

Assuming $c_{s}=\mathrm{const.}$, the set of Eqs. (\ref{Hdot}) and (\ref{c_s}) can be recast in the dimensionless form as
\begin{align}
 & 2 \tilde{H} \tilde{H}^\prime=-\left[3\Omega_{m_0} a^{-4}+\frac{a}{c_s} \tilde{H}^2 \tilde{\phi}'^2\right],
 \label{Heq}
 \\
 & a \tilde{H} \tilde{\phi} ^\prime+\left(\frac{1-c_s^2}{\tilde{f}_0}\right)^{1/2} \tilde{\phi}^2=0,
 \label{phieq}
\end{align}
where the prime represents the derivative with respect to the scale factor $a$. Here, we have used $X=\dot{\phi}^2/2$ and $\rho_m=\rho_{m_0}a^{-3}$. Also $\Omega_{m_0}=\rho_{m_0}/(3M_P^2H_0^2)$ is the dimensionless matter density parameter at the present. Furthermore, for the purpose of numerical computations, it is more convenient to define the following dimensionless variables
\begin{equation}
\begin{array}{lllll}
E\equiv\tilde{H}\equiv\frac{H}{H_{0}}, & & \tilde{\phi}\equiv\frac{\phi}{M_{P}}, & & \tilde{V}(\phi)\equiv\frac{V(\phi)}{H_{0}^{2}M_{P}^{2}},\\
\tilde{f}(\phi)\equiv f(\phi)H_{0}^{2}M_{P}^{2}, & & \tilde{f_{0}}\equiv\frac{f_{0}H_{0}^{2}}{M_{P}^{2}}.
\end{array}
\nonumber
\end{equation}

Notice that our DBI model described by Eqs. (\ref{Heq}) and (\ref{phieq}) has four free parameters including $\tilde{f}_0$, $c_s$, $\tilde{\phi}_0$, and $\Omega_{m_0}$. To investigate the background evolution and the growth of perturbations in our model, we consider two different cases. In the first case, we fix $\tilde{f}_0=0.04$ and consider different values of the sound speed $c_s = 0.05,\, 0.1,\, 0.9$. In the second one, we set $c_s=0.05$ and choose typical values for the warp parameter as $\tilde{f}_0 = 0.05,\, 0.07,\, 0.1$. In the both cases, we fix $\Omega_{m_0}=0.27$ and $\tilde{\phi}_0=0.14$ at the present time.

\begin{figure*}
\centering
\includegraphics[scale=0.46]{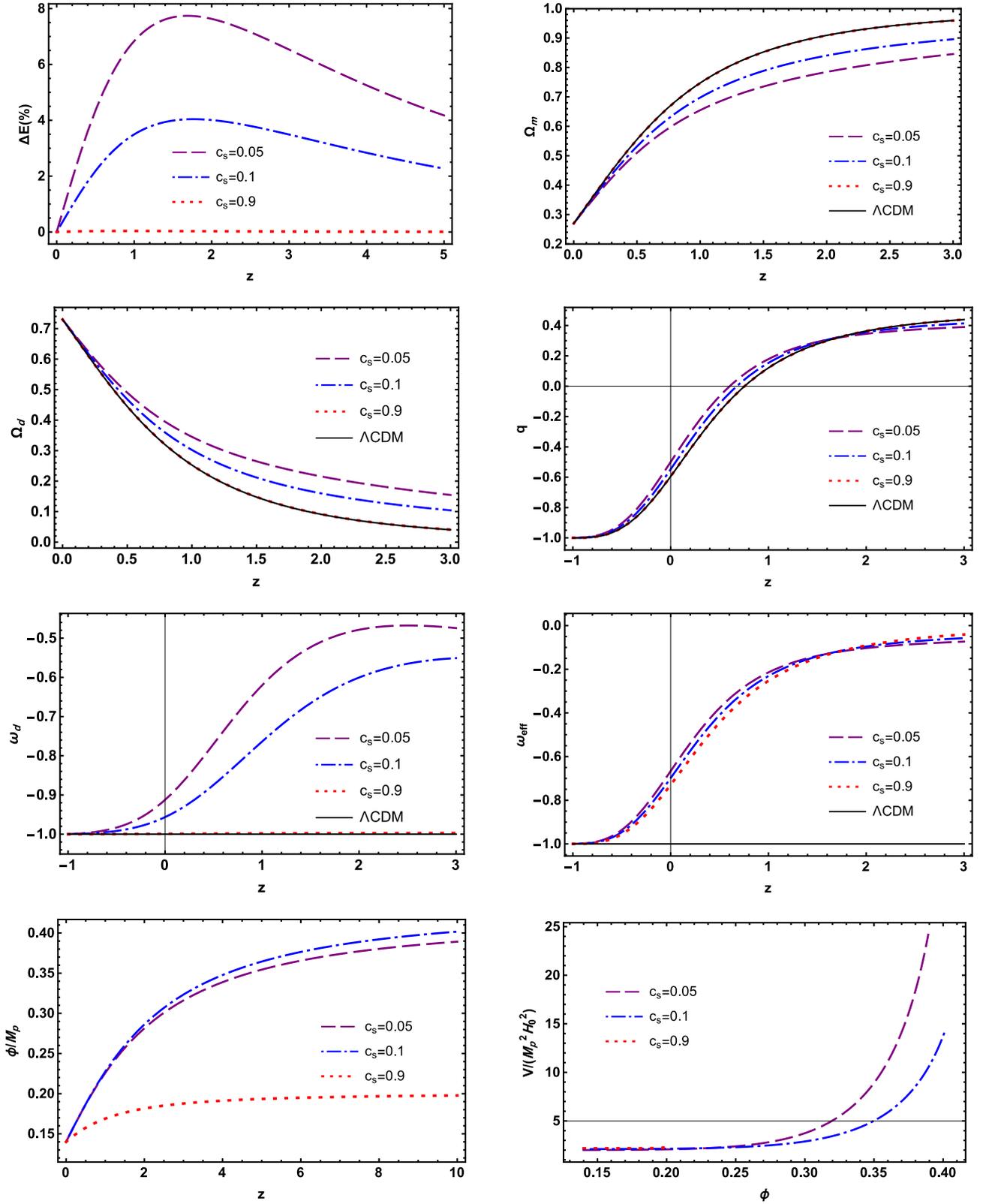}
\caption{\small{Variations of the relative deviation $ \Delta E(z)$ of the normalized Hubble parameter for the DBI models in comparison with the $\Lambda$CDM, the DM density parameter $\Omega_m$, the DE density parameter $\Omega_d$, the deceleration parameter $q$, the EoS parameter of DE $\omega_d$, the effective EoS parameter $\omega_{\rm eff}$, the DBI scalar field $\phi$, and the DBI potential $V(\phi)$. Auxiliary parameters are $\Omega_{m_0}=0.27$ and $\tilde{f}_0=0.04$.}}\label{cosmofig1}
\end{figure*}

\begin{figure*}
\centering
\includegraphics[scale=0.46]{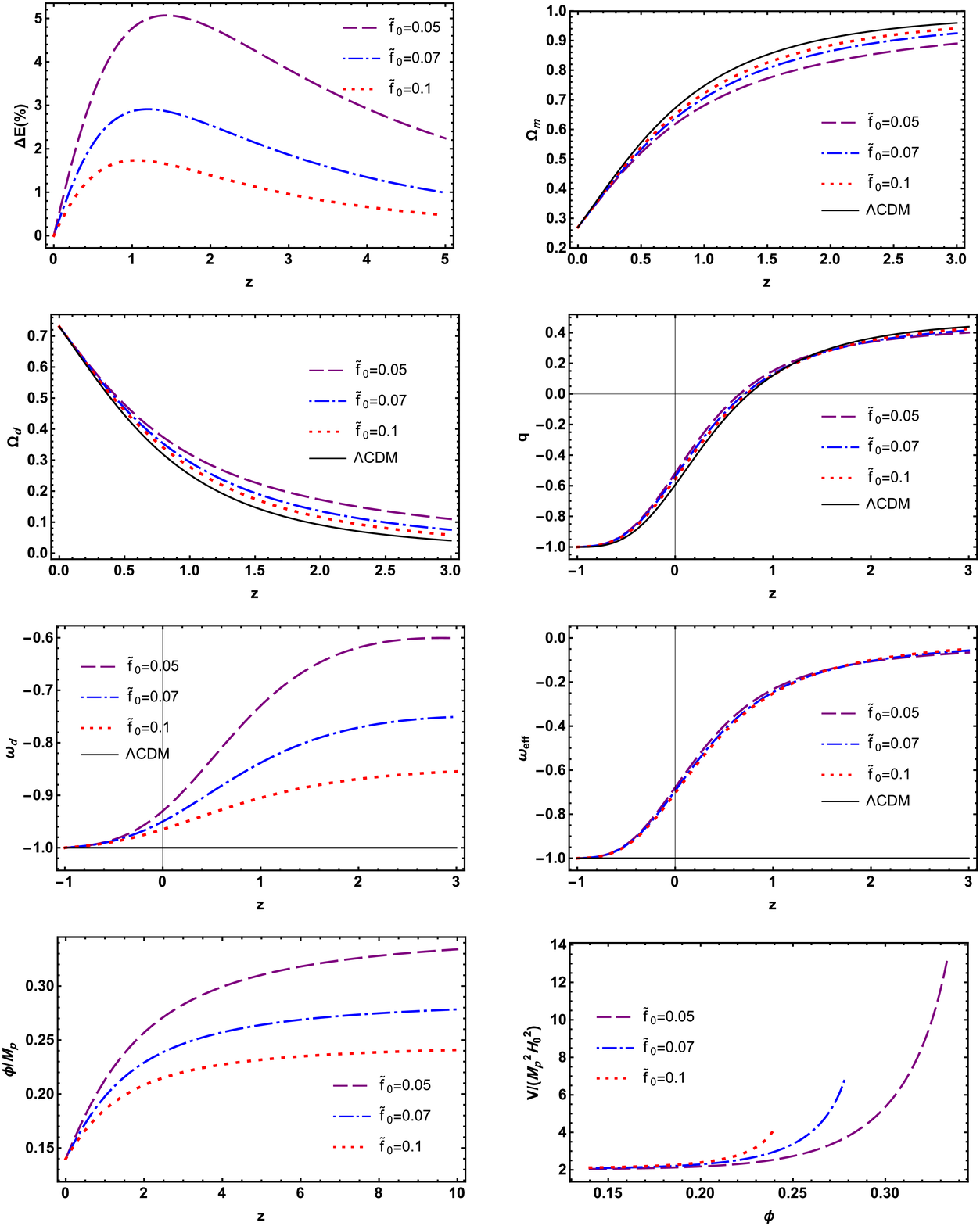}
\caption{\small{Same as Fig. \ref{cosmofig1}, but for $c_s=0.05$.}}\label{cosmofig2}
\end{figure*}

Now, we solve Eqs. (\ref{Heq}) and (\ref{phieq}) numerically with the initial conditions $\tilde{\phi}(a_0)=0.14$ and $\tilde{H}(a_0)=1$. With the help of numerical results obtained for $H$ and $\phi$, we can obtain the evolutionary behaviors of the DM and DE density parameters ($\Omega_m$, $\Omega_d$), the deceleration parameter $q=-1-\dot{H}/H^2$, the equation of state (EoS) parameter of DE $\omega_d\equiv p_d/\rho_d$, the effective EoS parameter $\omega_{\rm eff}=-1-\frac{2}{3}\frac{\dot{H}}{H^2}$, and the DBI potential $V(\phi)$. The results are presented in Figs. \ref{cosmofig1} and \ref{cosmofig2}. The figures show that (i) the DBI model with different $c_s<1$ and $\tilde{f}_0>0$ has $\Delta E=\Delta \tilde{H}=100\left[\frac{E_{{\rm DBI}}}{E_{\Lambda{\rm CDM}}}-1\right]>0$, which indicates that the Hubble parameter in our model is larger than one in the $\Lambda$CDM model. (ii) The density parameters $\Omega_d$ and $\Omega_m$ increase and decrease, respectively, as the redshift $z=\frac{1}{a}-1$ decreases. (iii) The deceleration parameter $q$ varies from an early matter-dominant epoch $(q = 0.5)$ to the de Sitter era ($q = -1$) in the late-time future, as expected. It also shows a transition from a cosmic deceleration ($q > 0$) to an accelerating phase ($q < 0$) in the near past. The transition occurs at the redshifts $z_t = (0.603, 0.675, 0.754$) for $c_s= (0.05$, $0.1$, $0.9$) with $\tilde{f}_0=0.04$ (see Fig. \ref{cosmofig1}) and $z_t=( 0.659$, $0.708$, $0.733)$ for $\tilde{f}_0=(0.05$, $0.07$, $0.1$) with $c_s=0.05$ (see Fig. \ref{cosmofig2}). The values of $z_t$ in our model are smaller than $z_t^{\Lambda \rm CDM}=0.755$ corresponding to the $\Lambda$CDM model. (iv) The EoS parameter of DE, $\omega_d$, in our model behaves like the quintessence regime,  i.e. $\omega_d>-1$. This result is similar to that obtained by \citet{Devi2011} for the tachyon DE model. (v) The effective EoS parameter, $\omega_{\rm eff}$, starts from an early matter-dominated regime ($\omega_{\rm eff}=0$) and it behaves like the $\Lambda$CDM model ($\omega_{\rm eff} \rightarrow -1$) in the late time ($z \rightarrow -1$). (vi) The DBI scalar field $\phi$ decreases with decreasing the redshift. (vii) The DBI potential for $c_s \rightarrow 1$ behaves like a nearly flat potential. Using the power-law fitting $V(\phi)\propto \phi^n$, we find that for $\tilde{f}_0=0.04$ the DBI potential behave as $V \propto {\phi}^4$, ${\phi}^2$, and $\rm{const.}$, for $c_s=0.05$, $0.1$, and $0.9$, respectively. Here it is important to note that in our DBI model, the slow-roll condition holds because the power-law potentials of our DBI models during the accelerating expansion period of the Universe ($q<0$) behave as flat potentials (see Figs. \ref{cosmofig1} and \ref{cosmofig2}, the last right panels). Now, according to \citet{Garriga1999} during slow-roll accelerating expansion, the Hubble parameter $H$ and the sound speed $c_s$ vary much slower than the scale factor $a$ of the Universe. Therefore, we have $\varepsilon\equiv -\frac{\dot H}{H^2}\ll 1$ and $s\equiv \frac{\dot{c_{s}}}{H c_{s}}\ll 1$. Consequently, in the slow-roll approximation, the assumption of constant sound speed, $c_s\simeq {\rm cte.}$, is valid. (viii) For smaller $c_s$ (or $\tilde{f}_0$), the differences between the DBI model and $\Lambda$CDM are more pronounced, while for $c_s \rightarrow 1$ the DBI model behaves like the $\Lambda$CDM one, as expected.

\section{Linear perturbation theory}
\label{section:linear}

In this section, we study the linear growth rate of density perturbations of non-relativistic DM and DBI dark energy. In the Pseudo-Newtonian (PN) formalism \citep{Hwang1997, Lima1997, Hwang2006}, the linear perturbation equations governing the evolution of non-relativistic dust matter $\delta_m\equiv \delta\rho_m/\rho_m$ and dark energy $\delta_d\equiv \delta\rho_d/\rho_d$ density contrasts are given by \citep{Abramo2009}
\begin{align}
 & \dot{\delta}_m+\frac{\theta_m}{a}=0,\label{dotdeltam}\\
 & \dot{\delta}_d+(1+\omega_d)\frac{\theta_d}{a}+3H\big(c_{\rm eff}^2-\omega_d\big)\delta_d =0,\label{dotdeltad}\\
 & \dot{\theta}_m+H\theta_m-\frac{k^2 \phi}{a}=0,\label{tetam}\\
 & \dot{\theta}_d+H \theta_d-\frac{k^2 c_{\rm eff}^2 \delta_d}{(1+\omega_d)a}-\frac{k^2 \phi}{a}=0,\label{tetad}
\end{align}
where $c_{\rm eff}^2 \equiv \delta p_d/\delta \rho_d$ is the effective sound speed of DE. Also $\theta_m\equiv \nabla\cdot \mathbf{v}_{m}$ and $\theta_d\equiv \nabla\cdot \mathbf{v}_{d}$ are the divergence of the comoving peculiar velocities for DM and DE, respectively. On the sub-horizon scales, the Poisson equation in the Fourier space takes the form \citep{Lima1997}
\begin{equation}
 \label{poisson}
 -\frac{k^2}{a^2}\phi =\frac{3}{2}H^2\big[\Omega_m \delta_m +\big(1+3~ c_{\rm eff}^2\big)\Omega_d \delta_d\big].
\end{equation}
In order to measure the evolution of DE and DM fluctuations, it is convenient to express Eqs. (\ref{dotdeltam})-(\ref{poisson}) in terms of the scale factor $a$, instead of the cosmic time $t$. We reach
\begin{align}
 & {\delta}^{\prime}_m+\frac{\tilde{\theta}_m}{a}=0 \;,\label{dadeltam}\\
 & {\delta}^{\prime}_d+\frac{3}{a}\big(c_{\rm eff}^2-\omega_d \big)\delta_d+(1+\omega_d)\frac{\tilde{\theta}_d}{a} =0 \;,\label{dadeltad}\\
 & \tilde{\theta}^{\prime}_m+\big(\frac{2}{a}+\frac{H^{\prime}}{H}\big)\tilde{\theta}_m+\frac{3}{2a}\left[\Omega_{m}\delta_{m}+ \big(1+3c_{\rm eff}^2 \big)\Omega_{d}\delta_{d}\right]=0 \;,\label{lithetam}\\
  & \tilde{\theta}^{\prime}_d+\big(\frac{2}{a}+\frac{H^{\prime}}{H}\big)\tilde{\theta}_d+\frac{3}{2a}\left[\Omega_{m}\delta_{m}+ \big(1+3c_{\rm eff}^2 \big)\Omega_{d}\delta_{d}\right]=0 \;,\label{lithetad}
\end{align}
where the prime denotes the derivative with respect to $a$, and $\tilde{\theta} \equiv \theta/H$ is the dimensionless form of the $\theta$ parameter. The set of above equations shows that the amount of DE clustering depends on the magnitude of its effective sound speed $c_{\rm eff}$. For $c_{\rm eff}=0$, the DE clusters in a similar manner to DM that we call it full clustering (FCL) DE, and $c_{\rm eff}=1$ is related to the case which DE is non-clustering (NCL), i.e. $\delta_d=0$. Note that the quantities $c_{\rm eff}$ and $c_s$ are different from each other. The adiabatic sound speed $c_s^2$ appears in the background cosmology (see Eqs. (\ref{Heq}) and (\ref{phieq})), while the effective sound speed $c_{\rm eff}$ only appears when any perturbation of DE is present.

We solve numerically the set of Eqs. (\ref{dadeltam})-(\ref{lithetad}) for both the full clustering ($c_{\rm eff}=0$, $\tilde{ \theta}_m=\tilde{\theta}_d$) and non-clustering ($c_{\rm eff}=1$, $\delta_d = 0=\tilde{\theta}_d$) DBI dark energy models, from an initial redshift $z_i=10^4$ at the equality epoch till the present time ($z=0$). We choose the initial conditions of an Einstein-de Sitter (EdS) Universe \citep{Batista2013, Pace2014}
\begin{align}
 & \delta_{m_i} = a_i=(1+z_i)^{-1},
 \label{deltami}
 \\
 &{\delta _d}_i=\left( \frac{1+\omega _{d_i}}{1-3\omega_{d_i}} \right) \delta_{m_i},
 \label{deltadi}
 \\
 &{\tilde{\theta}}_{m_i}=-\delta_{m_i},
\end{align}
which satisfy the set of Eqs. (\ref{dadeltam})-(\ref{lithetad}) in the matter-dominated era. With this choice the perturbations remain in the linear regime. Note that in general although during the matter-dominated regime the amplitude of DM perturbations behaves as $\delta_m = C a$ in which $C$ is a constant, and without any loss of generality one can set $C=1$.

\begin{figure*}
\begin{minipage}[b]{1\textwidth}
\subfigure[\label{deltama} ]{ \includegraphics[width=.48\textwidth]%
{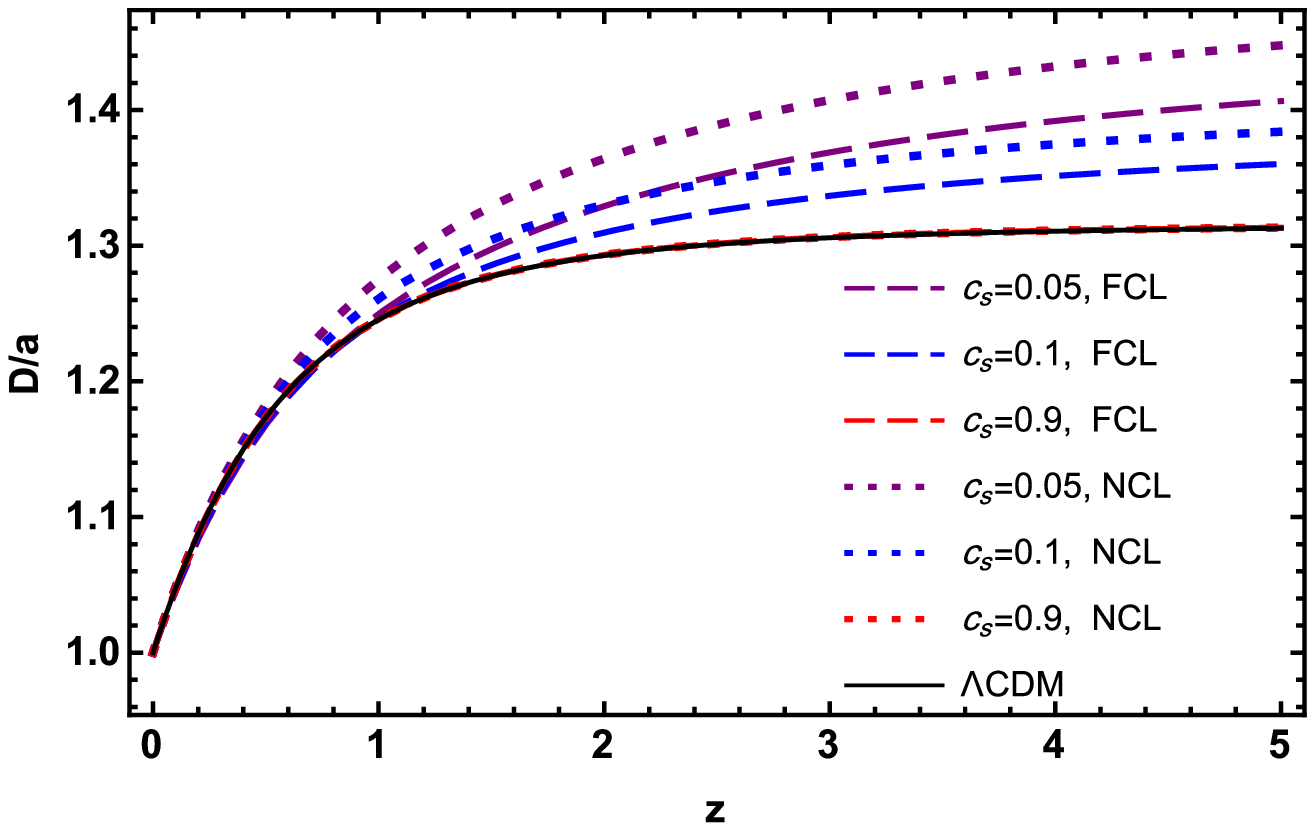}}\hspace{.1cm}
\subfigure[\label{deltamb}]{ \includegraphics[width=.48\textwidth]%
{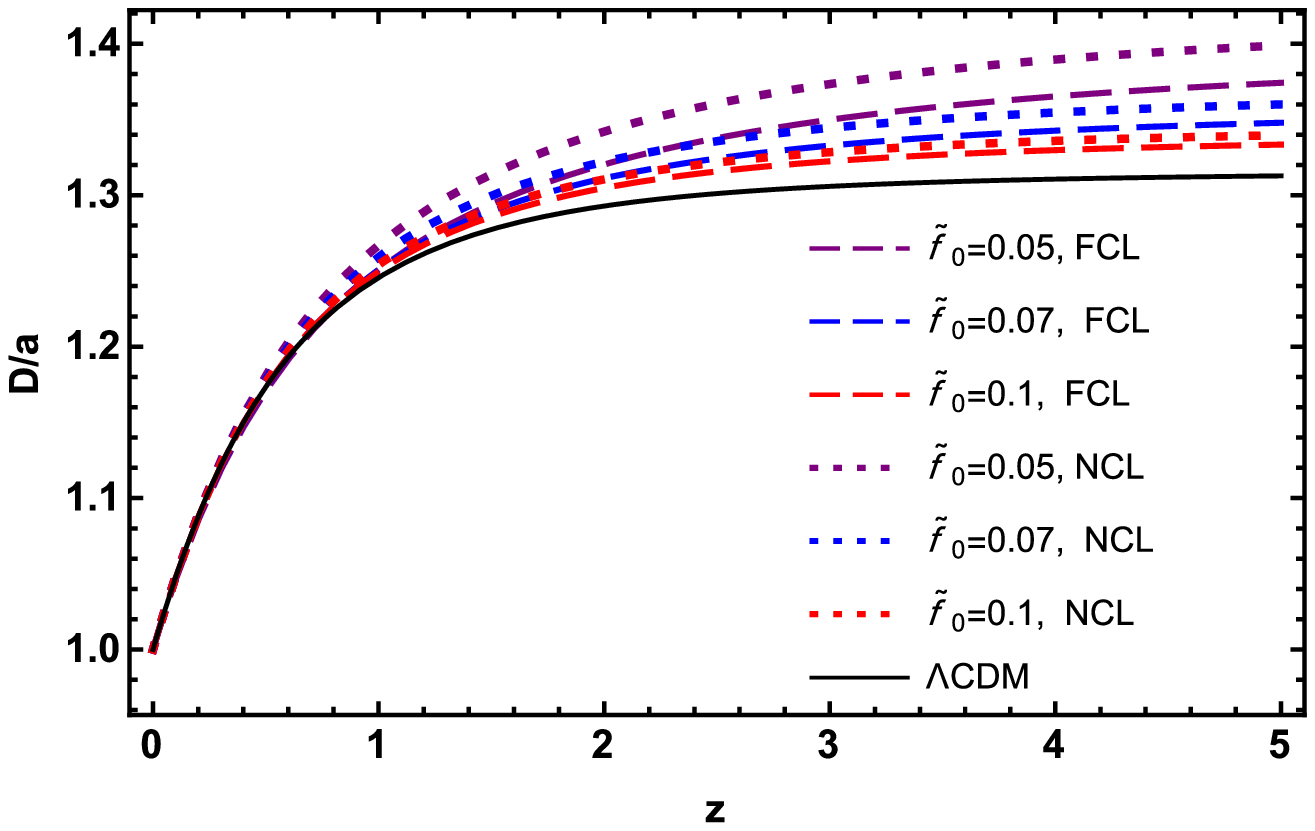}}
\end{minipage}
\caption{(a) Evolution of the growth function relative to its value in a pure
matter model $D/a$, in which $D=\delta_m/\delta_{m_0}$. Auxiliary parameters are $\Omega_{m_0}=0.27$ and $\tilde{f}_0=0.04$. (b) Same as Fig. \ref{deltama}, but for $c_s=0.05$.
  }\label{linear}
\end{figure*}

In Figs. \ref{deltama} and \ref{deltamb}, we plot the evolution of the growth factor, $D=\delta_m/\delta_{m_0}$, relative to its value in a pure matter model ($D=a$) versus the redshift $z$ for both the NCL and FCL DBI models with different sets of model parameters. The figures imply that (i) for a given $z$, the value of $D/a$ in the $\Lambda$CDM model is smaller than one in the DBI model. Also in the limit of $c_s \rightarrow 1$, the DBI model behaves like the $\Lambda$CDM one. (ii) Similarly to the $\Lambda$CDM Universe, in the DBI model, DE suppresses the growth of perturbations at low redshift, and this is due to the fact that at late times DE dominates the energy budget of the Universe. (iii) $D/a$ has a constant behavior at high redshifts, implying that at early times the effects of DE are negligible in all the models. (iv) The growth factor is smaller for clustering DBI model compared to the homogeneous one ($\delta_d=0$).

\section{Spherical collapse in the DBI model}
\label{section:spherical_collapse}

Here, we are interested in studying the non-linear evolution of DE and DM fluctuations in the framework of clustering DBI cosmology. The spherical collapse model (SCM) is the simplest analytical tool to study non-linear structure formation \citep{Gunn1972, Padmanabhan1993, Fosalba1998}. In the framework of SCM, a single spherically symmetric region with a peculiar expansion rate detaches from the homogeneous background. Due to self-gravity effects, the spherical overdense regions expand more slowly relative to the Hubble flow. This causes the density of the overdense sphere increases compared to the background fluid. Then, at the turnaround redshift, $z_{\rm ta}$, the spherical region reaches a maximum radius and completely decouples from the background fluid and begins to collapse independently. Eventually, at the virial redshift $z_{\rm vir}$, the collapsing sphere attains the steady state virial radius. The SCM is based on the crucial assumptions that the density of each component of fluid is always homogeneous in the spherical region (that means it follows the top-hat density profile), and that the velocity profile of each fluid keeps this homogeneity. Notice that in the presence of DE, not only the large-scale gravitational potentials grow slower because of the accelerated expansion of the Universe, but also the dynamical DE can cluster and form halos itself.

The equations governing the dynamical behavior of SCM are as follows \citep{Hu1998, Abramo2009, Pace2014}
\begin{align}
 & \dot{\delta}_j=-3 H (c_{{\rm eff}_j}^2-\omega_j) \delta_j-\big[1+\omega_j+\big(1+c_{{\rm eff}_j}^2\big)\delta_j\big] \frac{\theta}{a},
 \label{SC1}
 \\
 & \dot{\theta}=-H\theta -\frac{\theta^2}{3 a}-4 \pi G a \sum_j \rho_j \delta_j \big(1+3 c_{{\rm eff}_j}^2\big),
 \label{SC2}
\end{align}
where $\delta_j$, $c_{{\rm eff}_j}^2$ and $\omega_j$ are respectively the density contrast, the square of the effective sound speed and the EoS parameter of component $j$. Note that each fluid component $j$ obeys a separate equation of the type Eq. (\ref{SC1}), while Eq. (\ref{SC2}) stands alone and all fluids flow in the same way. This is true because in the SCM a top-hat density profile is used.

In the presence of full clustering DE ($c_{\rm eff}=0$), the set of equations (\ref{SC1}) and (\ref{SC2}) can be written in terms of scale factor as follows
\begin{align}
 &{\delta}^{\prime}_m+(1+\delta_m)\frac{\tilde{\theta}}{a}=0,
 \label{deltamSC}
 \\
 &{\delta}^{\prime}_d+(1+\omega_d+\delta_d)\frac{\tilde{\theta}}{a}-\frac{3}{a}~\omega_d\delta_d =0,
 \\
 &{\tilde{\theta}}^{\prime}+\left(\frac{2}{a}+\frac{H^\prime}{H}\right)\tilde{\theta}+\frac{{\tilde{\theta}}^2}{3 a}+\frac{3}{2a}\big(\Omega_m \delta_m +\Omega_d \delta_d\big)=0.
 \label{thetasc}
\end{align}
For the case of non-clustering DE ($c_{\rm eff}=1$), we only need to solve Eqs. (\ref{deltamSC}) and (\ref{thetasc}), setting $\delta_d=0$ in the latter, and in this case we have the usual SCM for DM \citep{Gunn1972, Padmanabhan1993, Percival2005}.

\begin{figure*}
\begin{minipage}[b]{1\textwidth}
\subfigure[\label{deltamnona} ]{ \includegraphics[width=0.48\textwidth]%
{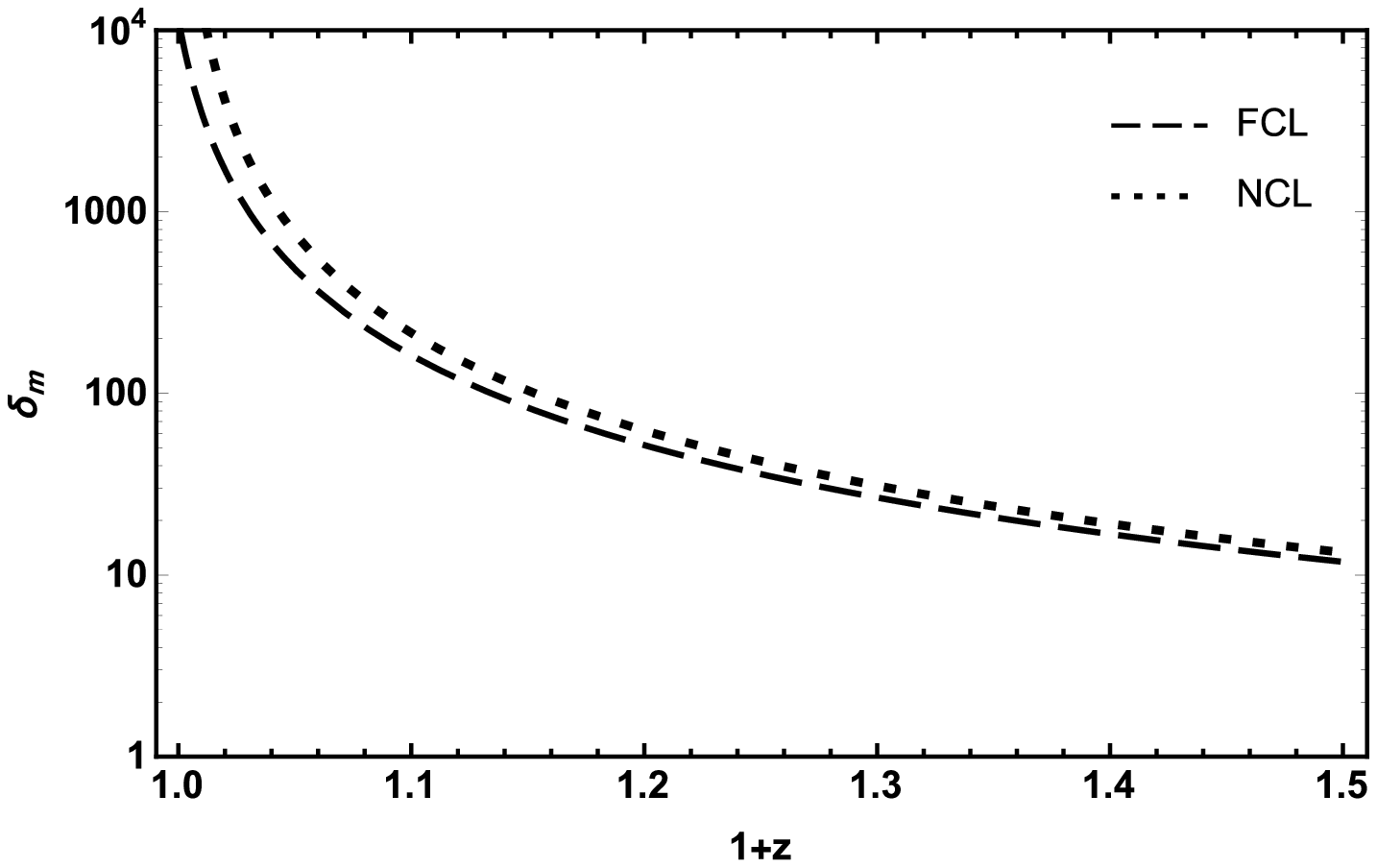}}\hspace{.1cm}
\subfigure[\label{deltanona}]{ \includegraphics[width=0.48\textwidth]%
{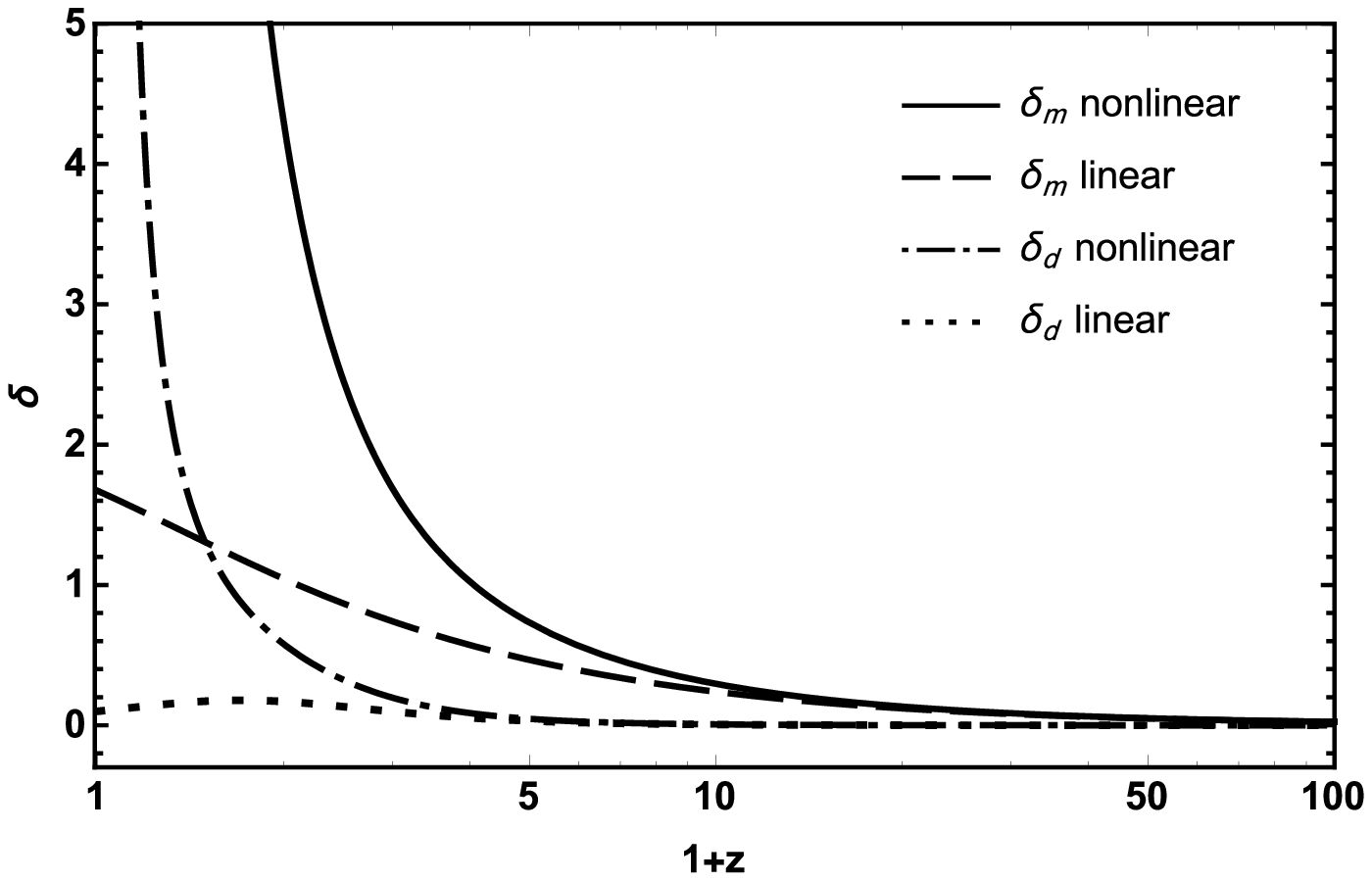}} \hspace{.1cm}
\end{minipage}
\caption{ (a) non-linear evolution of $\delta_m(z)$ with and without DE perturbation, (b) non-linear and linear evolutions of $\delta(z)$ for both the DM and DBI. Auxiliary parameters are $c_s=0.05$ and $\tilde{f}_0=0.04$.
  }\label{figdeltanon}
\end{figure*}

If we neglect the non-linear terms appeared in Eqs. (\ref{deltamSC})-(\ref{thetasc}), we reach the following set of equations for evolution of linear overdensities as
\begin{align}
 &{\delta}^{\prime}_m+\frac{\tilde{\theta}}{a}=0,
 \label{scdelta}
 \\
 &{\delta}^{\prime}_d+(1+\omega_d)\frac{\tilde{\theta}}{a}-\frac{3}{a}\omega_d\delta_d =0,
 \\
 &{\tilde{\theta}}^{\prime}+\left(\frac{2}{a}+\frac{H^\prime}{H}\right)\tilde{\theta}+\frac{3}{2a}\big[\Omega_m \delta_m +\Omega_d \delta_d\big]=0,
 \label{sctheta}
\end{align}
which are the same as Eqs. (\ref{dadeltam})-(\ref{lithetad}) in the PN formalism for the case of full clustering DBI dark energy ($c_{\rm eff}=0$, $\tilde{ \theta}_m=\tilde{\theta}_d \equiv \tilde{\theta}$). The non-linear evolution of $\delta_m$ with/without DBI perturbations with $c_s=0.05$ and $\tilde{f}_0=0.04$ are shown in Fig. \ref{deltamnona}. The initial conditions are chosen such that the spherical DM structure collapses at the present epoch, i.e. $\delta_m(z=0)\geq 10^7$. The figure shows that in the non-clustering DBI model, the DM structures at $z=0$ collapse earlier than those in the clustering DBI models. This means that in our DBI model which has a quintessence ($\omega_d>-1$) like behavior (see Figs. \ref{cosmofig1} and \ref{cosmofig2}), the DE overdensities inhibit the growth of DM perturbations. This is in well agreement with that obtained by \citet{Abramo2007} who showed that the inclusion of DE perturbations for the quintessence ($\omega_d>-1$, $\delta_d>0$) and phantom ($\omega_d<-1$, $\delta_d<0$) DE models, respectively, inhibit and enhance the growth of DM perturbations.

In Fig. \ref{deltanona}, we plot the linear/non-linear perturbations of both the DM and DBI dark energy with $c_s=0.05$ and $\tilde{f}_0=0.04$. We see that at early times both the linear and non-linear solutions behave very close to each other and at late times the non-linear solution grows very fast compared to the linear one. This also is in concordance with the result obtained by \citet{Abramo2007} for the clustering DE with non-phantom EoS parameter (i.e. $\omega_{d}>-1$).

\subsection{Spherical collapse parameters}
\label{subsection:spherical_collapse_parameters}

One of the main quantity characterizing the SCM is the critical density contrast or the linear overdensity parameter, $\delta_c$. It is defined as $\delta_c=\delta_{m \rm L}(z=z_c)$ in which $\delta_{m \rm L}$ is the linear matter density contrast computed from Eqs. (\ref{scdelta})-(\ref{sctheta}), with initial conditions such that the non-linear DM overdensity $\delta_m$ diverges at a given collapse redshift $z_c$  \citep{Pace2010, Pace2012, Pace2014}. Another important quantity in SCM is the virial overdensity defined as $\Delta_{\rm vir}=\zeta(x/y)^3$. Here $\zeta$ is the overdensity at the turn around epoch, $x$ is the scale factor normalized to the turn around scale factor and $y$ is the ratio between the  virialization radius and the turn around radius \citep{Wang1998}. In the Einstein-de Sitter (EdS) cosmology, one can easily show that $y=1/2$, $\zeta=5.6$ and $\Delta_{\rm vir}=178$ which are redshift-independent \citep{Meyer2012}. Note that in the presence of DE, the spherical collapse parameters can change in time. Also the virialization process depends on the DE evolution \citep{Lahav1991, Maor2005, Creminelli2010, Basse2012}.

\begin{figure*}
\centering
\includegraphics[scale=0.34]{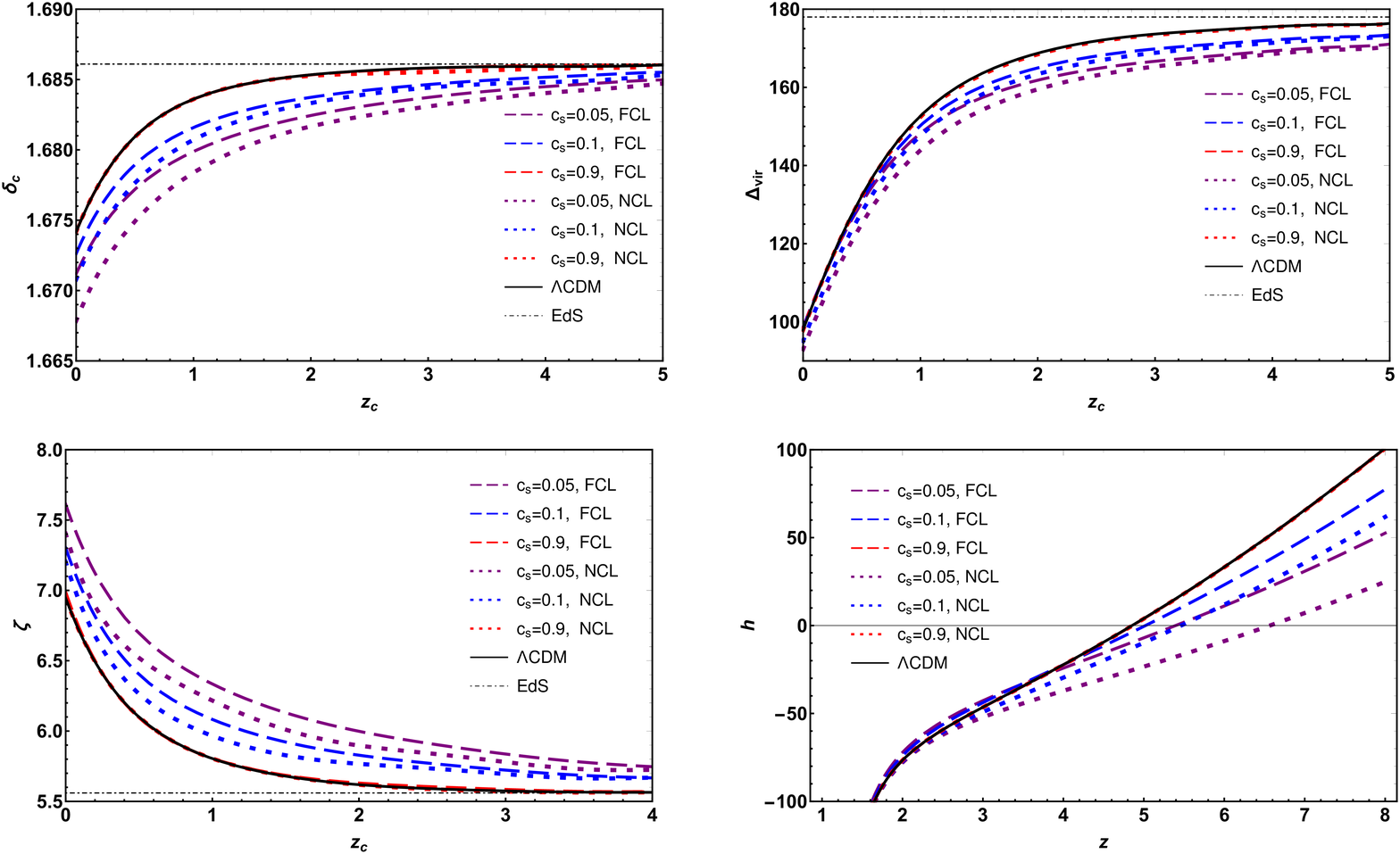}
\caption{\small{Evolutions of the critical density contrast $\delta_c$, the virial overdensity $\Delta_{\rm vir}$, the overdensity at the turn around $\zeta$, and the rate of expansion of collapsed region $h_{\rm ta}$. Auxiliary parameters are $\Omega_{m_0}=0.27$ and $\tilde{f}_0=0.04$.}}\label{figdeltac1}
\end{figure*}

\begin{figure*}
\centering
\includegraphics[scale=0.34]{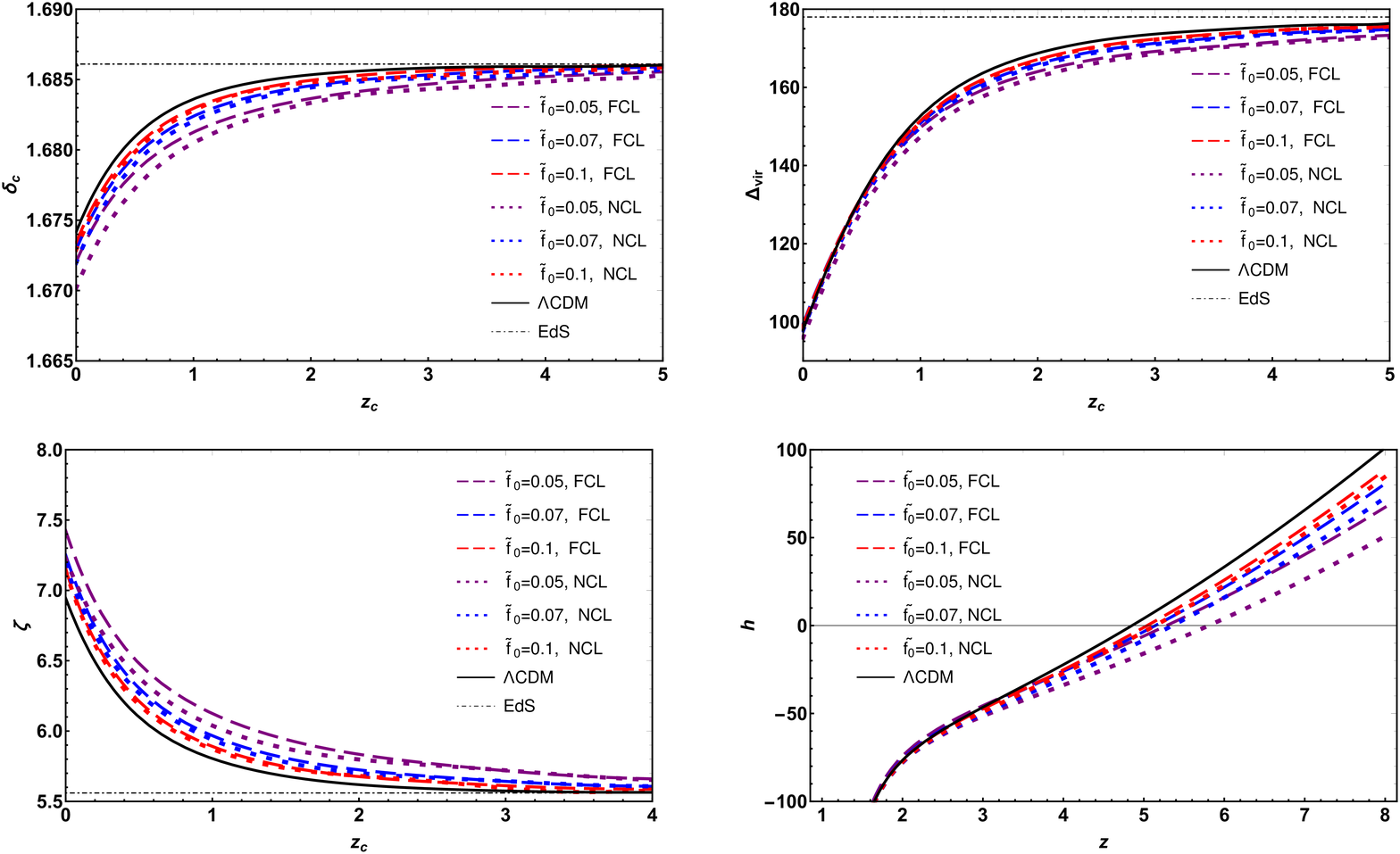}
\caption{\small{Same as Fig. \ref{figdeltac1}, but for $c_s=0.05$.}}\label{figdeltac2}
\end{figure*}

In Figs. \ref{figdeltac1} and \ref{figdeltac2}, the evolution of the linear overdensity $\delta_c(z_c)$, the virial overdensity $\Delta_{\rm vir}(z_c)$, the overdensity at the turn around $\zeta(z_c)$ and the rate of expansion of collapsed region, $h_{\rm ta}(z)= H (1+\theta/3a)$ \citep{Abramo2009JCAP} are presented for different sets of model parameters. The figures show that (i) for the case of FCL $(c_{\rm eff}=0)$, DBI perturbations clearly make $\delta_c$ closer to the $\Lambda$CDM compared to the corresponding non-clustering DBI models. This is in well agreement with what found by \citet{Pace2010, Batista2013, Pace2014, Malekjani2015}. (ii) At high redshifts, the linear overdensity tends to the fiducial value $\delta_c= 1.686$ in the EdS Universe. Note that \citet{Pace2017} have shown that the tendency of $\delta_c$ to the EdS limit at early times strongly depends on the value of the numerical infinity $\delta_{\infty}$ and on the choice of the initial time to start the integration of the equations, $a_i$. In our numerical calculations to satisfy the EdS limit, following \citet{Pace2017}, we set $\delta_{\infty}\geq 10^7$ and $a_i= 10^{-5}$. (iii) At lower redshifts, where DE dominates, $\delta_c$ decreases and deviates from the EdS limit. (iv) The virial overdensity in our DBI model at high enough redshift approaches the value in the EdS Universe, i.e. $\Delta_{\rm vir}=178$. Because the Universe is dominated by a pressureless dust matter and the effects of DE on structure formation are negligible. For the case of FCL, DE perturbations clearly make $\Delta_{\rm vir}$ closer to the result of the $\Lambda$CDM model compared to the corresponding NCL model. These implications are similar to what found by \citet{Del-Popolo2006, Del-Popolo2006-2, Pace2014, Malekjani2015}. (v) At high redshift, the values of overdensity at turn-around epoch $\zeta$ for DBI model asymptotically tends to the value in the EdS Universe $\zeta=5.55$. The value of $\zeta$ is larger for both the clustering and non-clustering DBI models, compared to the $\Lambda$CDM model. At lower redshifts, due to increasing the role of DE, $\zeta$ deviates from the EdS limit and the deviation for non-clustering DBI models is smaller than the clustering case. Note that the behaviors of $\Delta_{\rm vir}$ and $\zeta$ in our DBI model are similar to those obtained in \citet{Devi2011} for the tachyon DE model. (vi) The rate of expansion of collapsed region $h_{\rm ta}$ changes its sign from positive to negative value at the turn-around redshift. For $\tilde{f}_0=0.04$ with $c_s = (0.05, 0.1, 0.9)$, the transition occurs at the redshifts $z_{\rm ta} = ( 6.567, 5.463, 4.859)$ for non-clustering models and
$z_{\rm ta}=( 5.395, 5.021, 4.857)$ for clustering models. Also for $c_s=0.05$ with $\tilde{f}_0= (0.05, 0.07, 0.1)$, the transition occurs at the redshifts $z_{\rm ta} = ( 5.808,5.356,5.141)$ in non-clustering models and $z_{\rm ta}=(5.284,5.143,5.042)$ in clustering models. Note that for $\Lambda$CDM, the transition happens at $z_{\rm ta}=4.853$. We can see for non-clustering models, $h_{\rm ta}$ changes its sign faster. It means that for NCL case, turn-around epoch happens sooner compared to clustering DBI model. (vii) Notice that in all the figures, our DBI model in the limit of $c_s \rightarrow 1$ recovers the results of $\Lambda$CDM model.

\subsection{Mass function and halo number density}
\label{subsection:mass_function}

So far, we studied the impact of clustering DBI dark energy on the linear overdensity threshold $\delta_c$, the virial overdensity $\Delta_{\rm vir}$ and the overdensity at the turn around $\zeta$. Since we cannot directly observe the process of structure formation, it is convenient to determine a quantity closely related to the observations. This quantity is defined as the comoving number density of  virialized structures with masses in the certain range. Using a simple analytical method, \citet{Press1974} obtained the abundance of cold DM halos as a function of their mass and a Gaussian distribution function. In the Press-Schechter formalism, the comoving number density of virialized structures with masses in the range $M$ and $M+{\rm d}M$ at redshift $z$ is given by \citep{Press1974, Bond1991}
\begin{equation}
 \label{dnnoncl}
 \frac{{\rm d}n (M,z)}{{\rm d} M}=-\frac{\rho_{m_0}}{M}~\frac{{\rm d}{\rm ln}\sigma (M,z)}{{\rm d}M} f(\sigma),
\end{equation}
where $\rho_{m_0}$ is the background density of matter at the present time, $\sigma$ is the rms of the mass fluctuation in spheres of mass $M$, and $f(\sigma)=\sqrt{\frac{2}{\pi}}\frac{\delta_c}{\sigma}\exp{\big(-\frac{\delta_c^2}{2\sigma^2}\big)}$ is the standard mass function. Although the standard mass function works well in estimating the predicted number density of cold DM halos, it fails by predicting too many low-mass and too few high-mass objects \citep{Sheth1999, Sheth2002, Lima2004}. Hence, we use a more popular mass function introduced by Sheth and Tormen (ST) \citep{Sheth1999, Sheth2002} as follows
\begin{align}
 f_{\rm ST}(\sigma) = & A\sqrt{\frac{2 a}{\pi}}~ \left[1+\left(\frac{\sigma^2(M,z)}{a~ \delta^2_c(z)}\right)^p~\right]~ \frac{\delta_c(z)}{\sigma (M,z)}
 \nonumber
 \\
 & \times \exp\left(-\frac{a~ \delta_c^2}{2 \sigma^2(M,z)}\right),
\end{align}
where $A=0.3222$, $a=0.707$ and $p=0.3$. Following \citet{Abramo2007}, the quantity $\sigma(M,z)$ can be related to its present value as $\sigma(M,z)=D(z) \sigma_M$, where $D(z)=\delta_m(z)/\delta_m(z=0)$ is the linear growth function. Also, $\sigma_M^2$ is the variance of smoothed linear matter density contrast defined as
\begin{equation}
 \label{sigmaR}
 \sigma_M^2= \int_{0}^{\infty} \frac{d k}{k} \frac{k^3}{2 \pi ^2} P(k) W^2(kR) ~,
\end{equation}
where $R$ is the scale enclosing the mass $M=(4\pi/3)R^3 {\rho}_{m_0}$, and $W(kR)=\frac{3}{(kR)^3}\big(\sin(kR)-kR \cos(kR)\big)$ is a top-hat window function to carry out the smoothing. Also, $P(k)$ is the matter power spectrum of density fluctuations given by \citep{Liddle1993, Liddle1996}
\begin{equation}\label{powerspectrum}
 \frac{k^3 }{2 \pi ^2}  P(k)= \delta_{H_0}^2 \left(\frac{c k}{H_0} \right)^{n_s+3} ~T^2(k)~,
\end{equation}
where $n_s=0.968$ \citep{Planck2016} is the spectral index of primordial perturbation, $c$ is the speed of light, and $\delta_{H_0}$ is the present day normalization of the power spectrum. Besides, $T(k)$ is the transfer function which depends on cosmological parameters and the nature of the matter in the universe. Here, we use the Bardeen-Bond-Kaiser-Szalay (BBKS) transfer function which is given by \citep{Bardeen1986},
\begin{align}\label{transfer function}
  T(x)=&\frac{\ln (1+ 2.34 x)}{2.34 x}\nonumber
  \\& \times
  \left[1+3.89 x+(16.1 x)^2+(5.46 x)^3+(6.71 x)^4 \right]^{-1/4},
\end{align}
with $x\equiv k/h \Gamma$ where $\Gamma$ is the shape parameter defined as \citep{Sugiyama1995}
\begin{equation}
 \label{Gamma}
 \Gamma=\Omega_{m_0} h \exp \left( -\Omega_B - \Omega_B/\Omega_{m_0} \right).
\end{equation}
Here, $\Omega_B$ is the baryon density parameter, which we take it as $0.016 h^{-2}$ \citep{Copi1995, Copi1995-2}.
Note that one may use the fitting formulae of \citet{Bunn1997} to normalize the power spectrum to the COBE Differential Microwave Radiometer measurment. But here we normalize the power spectrum to the same value today, according to, $\sigma_{8}=\sigma_{8,\Lambda}\frac{\delta_c(z=0)}{\delta_{c,\Lambda}(z=0)}$, where $\sigma_{8,\Lambda}=0.8$ is used to normalize the matter power spectrum of $\Lambda$CDM \citep{Planck2016}.

\begin{figure*}
\begin{minipage}[b]{1\textwidth}
\subfigure[\label{figep1}]{ \includegraphics[width=.48\textwidth]%
{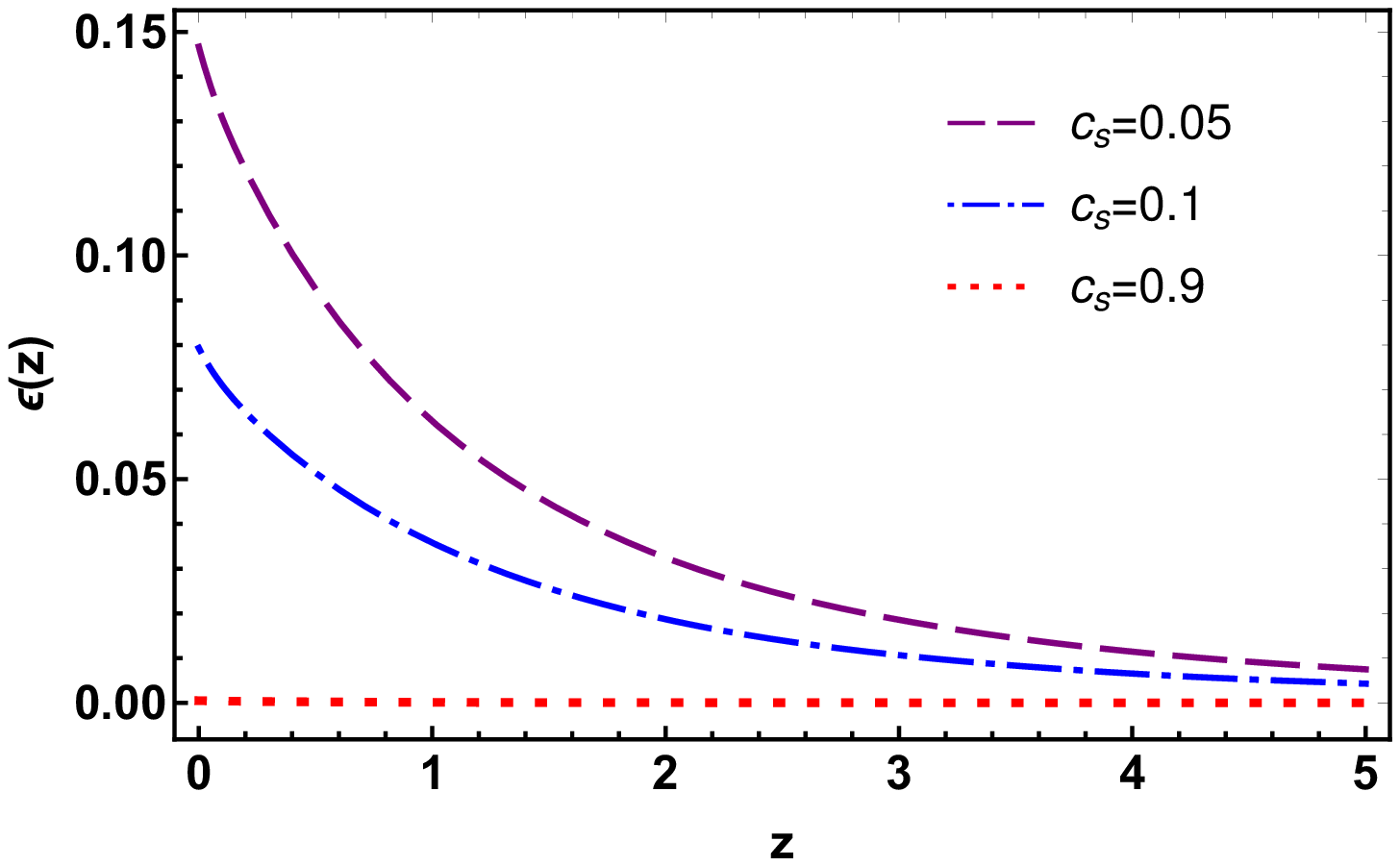}}\hspace{.1cm}
\subfigure[\label{figep2}]{ \includegraphics[width=.48\textwidth]%
{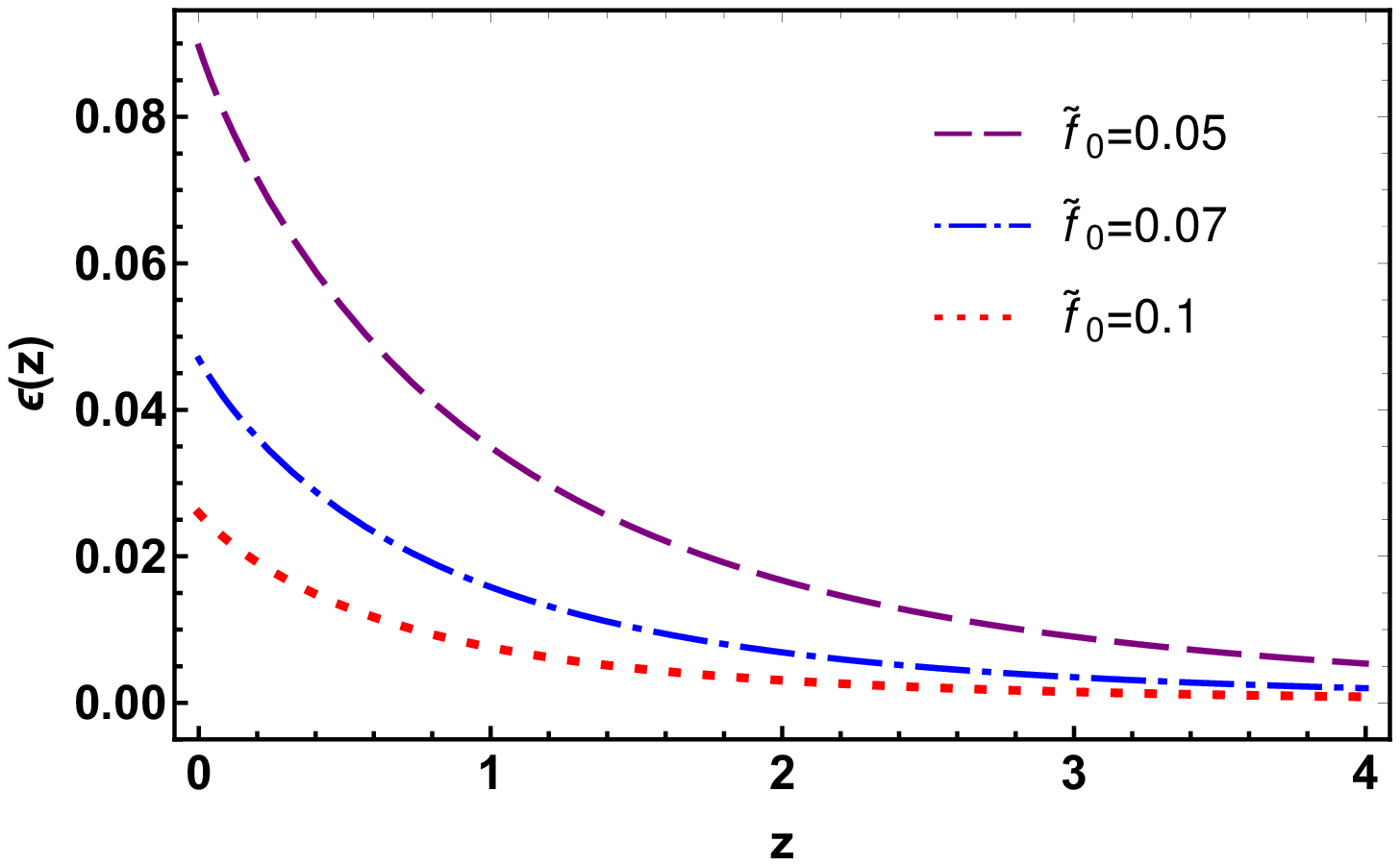}}
\end{minipage}
\caption{ The ratio of DBI dark energy mass to DM mass for (a) $\tilde{f}_0=0.04$ and (b) $c_s=0.05$.
  }\label{figepsilon}
\end{figure*}

In DE clustering scenario, the perturbations of DE can contribute to the halo mass, thus we must care about its contribution and we should take it into account \citep{Creminelli2010, Basse2011, Batista2013, Pace2014, Malekjani2015}. The fraction of DE mass to the mass of DM is given by the quantity $\epsilon(z)=M_d/M_m$, where in the case of full clustering DE and top-hat density profile, we have
\begin{equation}
 \label{epsilon}
 \epsilon(z)=\frac{\Omega_{d}(z)}{\Omega_m(z)}\left(\frac{\delta_{d}}{1+\delta_m}\right).
\end{equation}
In Fig. \ref{figepsilon}, we show the evolution of $\epsilon(z)$  on the base of Eq. (\ref{epsilon}) for our DBI model with different $c_s$ and $\tilde{f}_0$. The figure illustrates that (i) at earlier times, $\epsilon$ approaches zero. This indicates that the contribution of DE mass to the total mass of halos at high redshifts is negligible. (ii) In the left (right) panel, smaller $c_s$ (smaller $\tilde{f}_0$) gives a higher contribution of DE to the total mass of halos. (iii) When $c_s \rightarrow 1$, we can see the quantity of $\epsilon(z)$ tends to zero due to $\delta_{d} \rightarrow 0$. (iv) For all DBI models studied in this work, $\epsilon(z)$ is positive. This is because of this fact that for clustering DE models, from the initial condition (\ref{deltadi}) the evolution of $\delta_d$ at early times depends on the EoS parameter $\omega_d$ \citep{Abramo2007, Pace2014}. Since our DBI models behave like the quintessence DE ($\omega_d>-1$) hence we have $\delta_d>0$ and consequently from Eq. (\ref{epsilon}) we obtain $\epsilon>0$.

In the presence of DE contribution to the halo mass which appears in the parameter $\epsilon(z)$, Eq. (\ref{dnnoncl}) should be corrected as follows \citep{Batista2013, Pace2014-2}
\begin{equation}
 \label{dncl}
 \frac{{\rm d}n (M,z)}{{\rm d} M}=\frac{\rho_{m_0}}{M (1-\epsilon)}\frac{{\rm d} {\rm ln} \sigma (M,z)}{{\rm d}M} f(\sigma),
 \end{equation}
where the halo mass is changed by $M \rightarrow M(1-\epsilon)$. It should be noted that the clustering of the DE component can also change the mass function $f(\sigma)$ by changing the quantities $\delta_c$ and $\sigma(M,z)$. For the homogeneous and clustering DBI dark energy models, respectively, we use Eqs. (\ref{dnnoncl}) and (\ref{dncl}) to compute the number density of objects above a given mass at fixed redshift as $n(>M)=\int_{M}^{\infty} \frac{{\rm d} n}{{\rm d}M'}~ {\rm d}M'$.

\begin{figure*}
\centering
\includegraphics[scale=0.34]{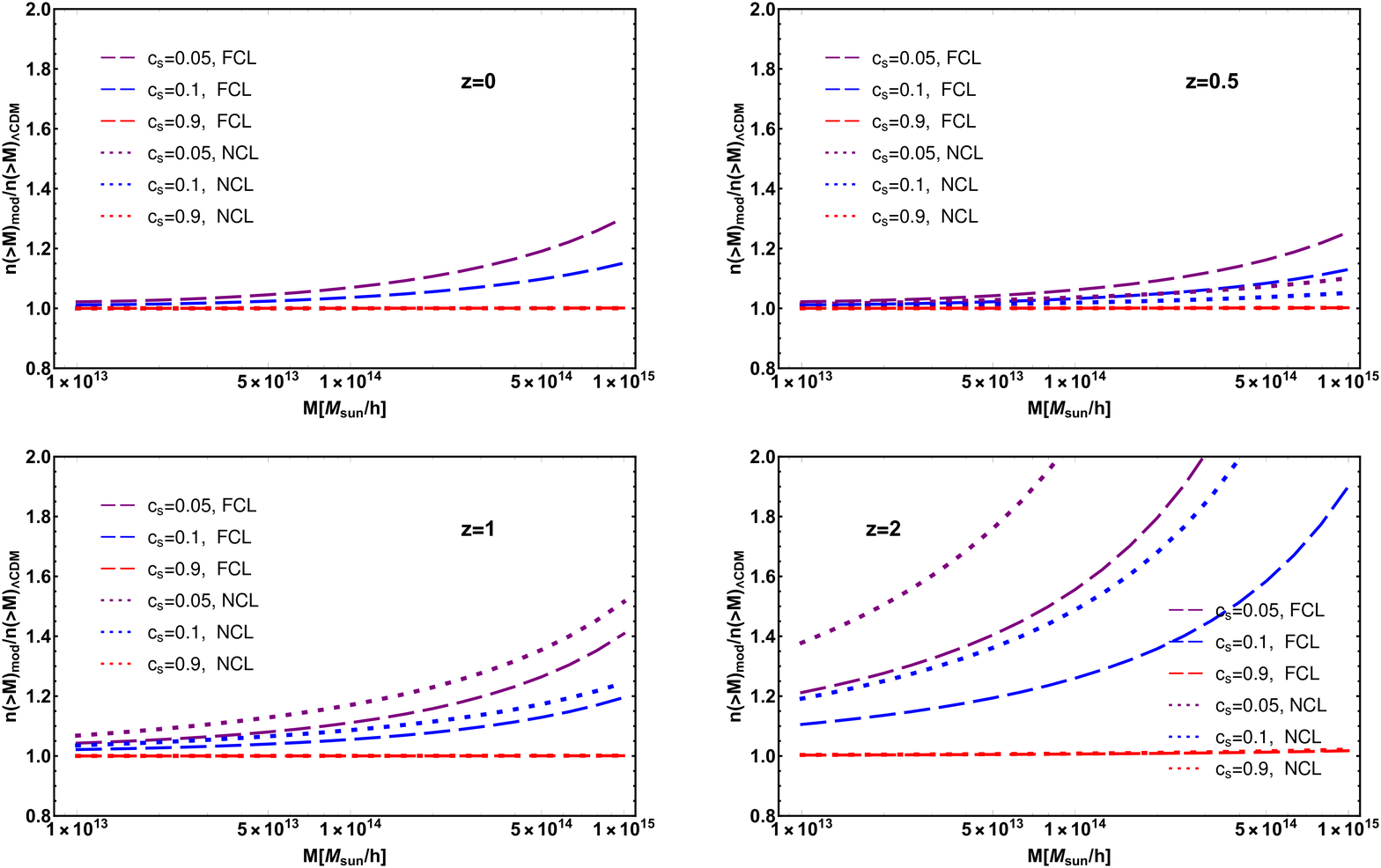}
\caption{\small{The relative number of halo objects above a given mass M at the redshifts $z=0, 0.5, 1, 2$ for $\tilde{f}_0=0.04$ and different $c_s$.}}
\label{mass-fig1}
\end{figure*}

In Fig. \ref{mass-fig1}, we show the relative number density of halos above a given mass at different fixed redshifts $z=0, 0.5, 1, 2$ for both the NCL and FCL DBI models. We see that at $z=0$ (or $z=0.5$), the results of all non-clustering DBI models roughly coincide with (or close to) the $\Lambda$CDM model. But for clustering DBI models with smaller values of $c_s$, the number of objects is more than $\Lambda$CDM model. At redshifts $z=1$ and $z=2$, the number of virialized halos estimated in clustering DBI models is lower than that for homogeneous DBI models. In general, the differences between clustering and homogeneous DBI models with the $\Lambda$CDM model are more pronounced at high redshifts and in the high-mass tail of the mass function. This is because of this fact that in the Sheth and Tormen mass function \citep{Sheth1999, Sheth2002}, the linear overdensity parameter $\delta_c$ plays an important role. A small variation of $\delta_c$ has a huge effect on the high-mass tail of the mass function \citep{Batista2013, Pace2014, Heneka2017}.

\begin{figure*}
\centering
\includegraphics[scale=0.34]{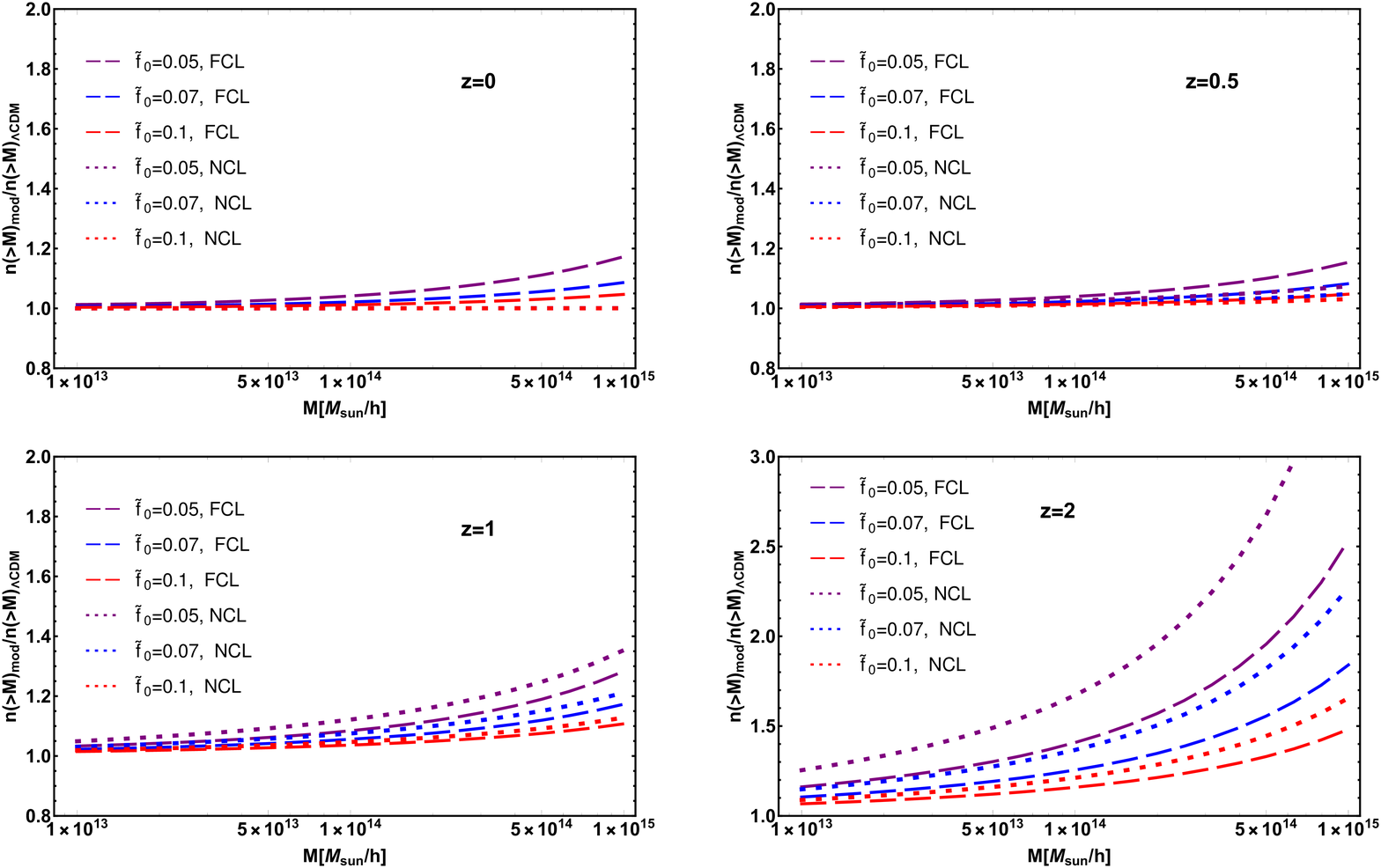}
\caption{\small{Same as Fig. \ref{mass-fig1}, but for $c_s=0.05$ and different $\tilde{f}_0$.}}\label{mass-fig2}
\end{figure*}

Figure \ref{mass-fig2} is the same as Fig. \ref{mass-fig1} but for different values of $\tilde{f}_0$. The figure clears that for clustering and non-clustering DBI models, the number density is almost equal to or larger than the $\Lambda$CDM model. Also similar to Fig. \ref{mass-fig1}, the predicted number of halos in clustering models at lower and higher redshifts, respectively, is more and less than homogeneous DBI models.

\section{Conclusions}
\label{section:conclusions}
Here, we studied both the linear and non-linear growth of DM and DE perturbations in the context of DBI non-canonical scalar field. We considered a DBI model with the AdS warp factor $f(\phi)=f_0\, \phi^{-4}$ and constant sound speed $c_s$. For the background cosmology, we assumed a spatially flat FRW Universe containing the pressureless DM and DBI dark energy. Then, we investigated the evolution of the background quantities including the Hubble parameter $H$, the density parameters ($\Omega_m$, $\Omega_d$), the deceleration parameter $q$, the DBI and effective EoS parameters ($\omega_d$, $\omega_{\rm eff}$) and the DBI scalar field potential $V(\phi)$. Our results show that (i) for smaller $c_s$ (or $f_0$), the background quantities in our DBI model deviate more than those in the $\Lambda$CDM model. (ii) $\omega_d$ behaves like quintessence DE, i.e. $\omega_d>-1$. (iii) $\omega_{\rm eff}$ and $q$ vary from matter dominated Universe $(\omega_{\rm eff}=0,q=0.5)$ and approach the de Sitter Universe $(\omega_{\rm eff}=-1,q=-1)$ at late times, as expected. Besides, $q$ shows a transition from decelerating ($q>0$) to accelerating ($q<0$) Universe at redshifts close to the $\Lambda$CDM model. (iv) For a given $f_0$ and different values of $c_s$, the DBI potential behaves like the power-law one $V(\phi)\propto \phi^n$.

In the linear regime of perturbations based on the PN formalism, we obtained the growth factor, $D=\delta_m/\delta_{m_0}$, relative to its value in a pure matter model ($D=a$). We found that the growth rate of DM in DBI models with smaller $c_s$ (or $\tilde{f}_0$) has larger deviations from the $\Lambda$CDM model and at the same time these deviations are smaller for clustering models compared to the homogeneous ones.

To study the growth of DM and DBI overdensities, we used the SCM and calculated the linear overdensity $\delta_c(z_c)$, the virial overdensity $\Delta_{\rm vir}(z_c)$, the overdensity at the turn around $\zeta(z_c)$ and the rate of expansion of collapsed region $h_{\rm ta}(z)$. Our results are summarized as follows. (i) For all non-clustering DBI models, the linear overdensity $\delta_c$ has bigger deviations from $\Lambda$CDM model in comparison with clustering DBI models. Also this deviation for DBI models with smaller $c_s$ (or $\tilde{f}_0$) is larger than the $\Lambda$CDM model. (ii) The virial overdensity $\Delta_{\rm vir}$ and the overdensity at the turn around $\zeta$ approach $178$ and $5.55$ in high enough redshifts, respectively, which are the same values obtained in the EdS cosmology. This is to be expected, because the impact of DE on the early evolution of the Universe is negligible. In addition, the values of $\Delta_{\rm vir}(z_c)$ and $\zeta(z_c)$ for the smaller $c_s$ (or $\tilde{f}_0$) deviate more than the $\Lambda$CDM model. (iii) For larger values of $c_s $ (or $\tilde{f}_0$), the rate of expansion of collapsed region $h_{\rm ta}(z)$ is almost similar to the $\Lambda$CDM model. Also for the homogeneous DBI models, $h_{\rm ta}$ changes its sign at higher redshifts compared to the clustering ones. This means that for the non-clustering models, the turn-around epochs occur sooner than ones in the clustering DBI models.

Finally, with the help of spherical collapse parameters, we calculated the ratio of DE mass to DM one $\epsilon(z)=M_d/M_m$ and the relative number density of objects above a given mass $\frac{n(>M)_{\rm DBI}}{n(>M)_{\Lambda \rm CDM}}$. We found that in the case of $z=0$, the number density of halos in non-clustering DBI models is very close to the $\Lambda$CDM model, but in clustering DBI models with smaller $c_s$ it is bigger than one in the $\Lambda$CDM. At redshifts $z=1$ and $z=2$, the number of virialized halos estimated in clustering models is lower than one in the homogeneous DBI models. In summary, the differences between clustering and homogeneous DBI models are more pronounced at high redshifts and in the high-mass tail of the mass function.

\section*{Acknowledgments}

The authors thank the anonymous referee for very valuable comments. The work of S. Asadzadeh has been supported financially by Research Institute for Astronomy and Astrophysics of Maragha (RIAAM) under research project No. 1/5440-18.












\bsp	
\label{lastpage}
\end{document}